\documentclass[amsmath,amssymb,aps,pra,twocolumn,superscriptaddress]{revtex4-1}

\usepackage[T1]{fontenc}
\usepackage{amsfonts}
\usepackage{braket}
\usepackage{graphicx}
\usepackage{float}
\usepackage{placeins}
\usepackage{hyperref}
\hypersetup{hidelinks}

\newcommand{\Tr}{\operatorname{Tr}}
\newcommand{\tr}{\operatorname{Tr}}
\newcommand{\dd}{\mathrm{d}}

\begin{document}

\title{Locally optimized variational evolution for quantum many-body systems}

\author{Carolin Wille}
\affiliation{London Centre for Nanotechnology, University College London, Gordon St., London, WC1H 0AH, United Kingdom}
\author{Maximilian Marvell}
\affiliation{London Centre for Nanotechnology, University College London, Gordon St., London, WC1H 0AH, United Kingdom}
\author{Lauren Stewart}
\affiliation{London Centre for Nanotechnology, University College London, Gordon St., London, WC1H 0AH, United Kingdom}
\author{Maximillian Murphy}
\affiliation{London Centre for Nanotechnology, University College London, Gordon St., London, WC1H 0AH, United Kingdom}
\author{Vinul Wimalaweera}
\affiliation{London Centre for Nanotechnology, University College London, Gordon St., London, WC1H 0AH, United Kingdom}
\author{Lesley Gover}
\affiliation{London Centre for Nanotechnology, University College London, Gordon St., London, WC1H 0AH, United Kingdom}
\author{Andrew~G. Green}
\affiliation{London Centre for Nanotechnology, University College London, Gordon St., London, WC1H 0AH, United Kingdom}

\date{\today}

\begin{abstract}

Quantum many-body dynamics is difficult to simulate because the global wavefunction develops extensive complexity, whereas local observables may retain only an intrinsic, system-size-independent complexity. This distinction is especially relevant in thermalizing systems, where local subsystems lose microscopic information and relax towards equilibrium. We introduce a variational time-evolution principle that exploits this separation by replacing global-state fidelity with a cost function defined on local reduced density matrices. The resulting locally optimized matrix-product-state (LOMPS) algorithm retains coherent short-time dynamics while using the simplification generated by thermalization at later times, and admits closed-form equations of motion analogous to the time-dependent variational principle. The same principle yields a quantum--classical algorithm in which a quantum processor evaluates the local cost and a noise-robust classical procedure performs the optimization. Proof-of-principle implementations on Quantinuum H2 and IBM Heron processors recover the characteristic local dynamics.

\end{abstract}

\maketitle

\begin{figure*}[!t]
  \centering

  \includegraphics[width=\textwidth]{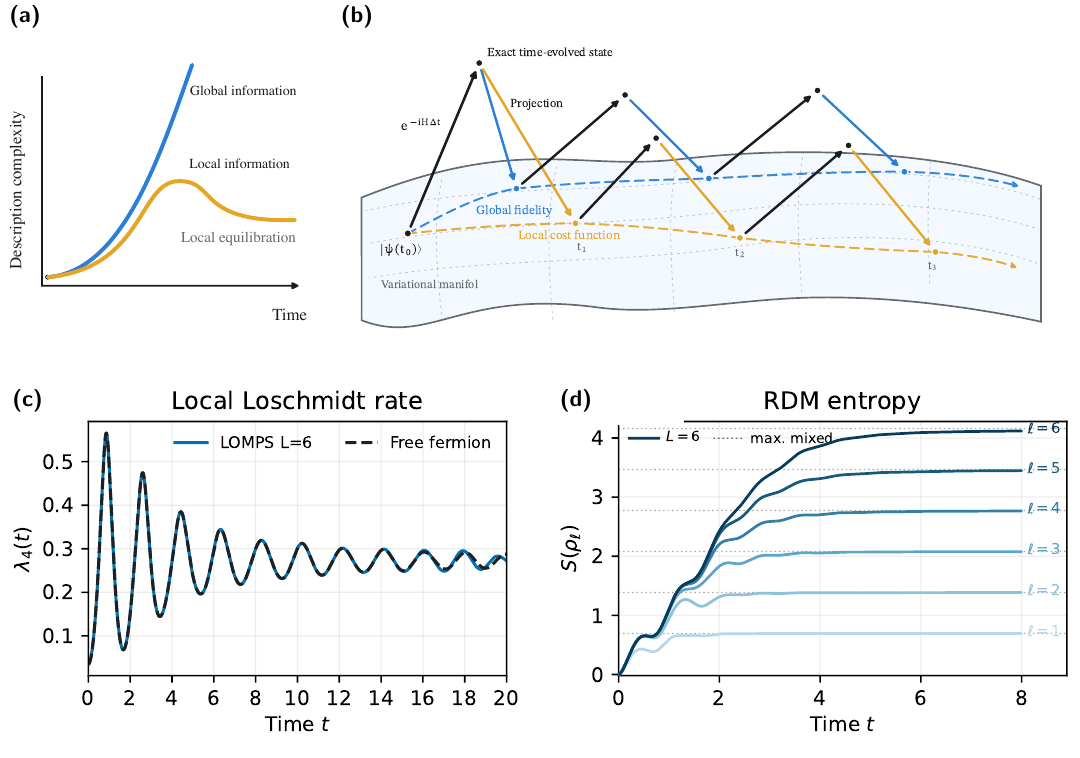}
  \caption{\textbf{(a)} \textbf{Evolution of local vs global description complexity.}
  The curves presented here are symbolic. For a concrete example of local description complexity dynamics see App.~\ref{app:osee}.
  \textbf{(b)} \textbf{Variational time evolution based on global versus local
  optimization.} The Hamiltonian dynamics takes the system off the variational manifold. Cost-function optimization projects the state back onto the variational manifold. Iterating this procedure defines a trajectory that depends on the chosen cost function. \textbf{(c)} \textbf{Early-time coherence} as demonstrated by the agreement of the four-site local Loschmidt rate $\lambda_4(t)$ (cf. Eq.~\eqref{eq:local-loschmidt}) obtained by our local variational evolution algorithm (LOMPS) compared to a quasi-exact free-fermion calculation for a quench of the integrable transverse-field Ising model.
  \textbf{(d)} \textbf{Late-time thermalization.} In the thermalising non-integrable
  tilted-field Ising quench local reduced density matrices on $\ell$ sites thermalize to the maximally mixed state. Solid curves show the $L=6$ LOMPS result compared to the expected maximally mixed values
  (dotted) at late times.}
  \label{fig:main_figure}
\end{figure*}

\section{Introduction}
Quantum many-body time evolution is difficult to simulate classically because the rapid growth of entanglement makes the global wave function increasingly costly to represent. This difficulty is a central motivation for quantum computers, which are expected to offer an extensive advantage over classical computers. However, physically accessible observables are local in nature and their complexity remains bounded even as the global wave function becomes increasingly complex as illustrated in Fig.~\ref{fig:main_figure}(a). Because of this the quantum advantage for simulating local observables should not be expected to be extensive, but instead governed by a system-dependent intrinsic complexity of local evolution~\cite{LocallyAccurateTN,HamiltonianQCA}.

The reduced complexity of local subsystems is especially prominent for thermalizing systems for which the eigenstate thermalization hypothesis (ETH) holds. Here, the structure of local observables is heavily constrained. As distant features of the wave function dephase, local subsystems lose memory of microscopic details and relax towards equilibrium values determined by only a few parameters. The difficulty in accurately capturing the dynamics of local observables thus depends on intensive features like the presence of ergodicity breaking or ergodicity delaying features and the emerging local relaxation time scales. A classical algorithm that targets the variational evolution of local observables should scale with this intrinsic complexity, rather than with the system size. 

In the following we present a concrete variational algorithm that aims to circumvent the complexity barrier when time evolving local observables by explicitly optimizing a \emph{local} cost function rather than a global state fidelity. The algorithm performs effective time evolution on a variational manifold (see Fig.~\ref{fig:main_figure}(b)) by alternating time evolution, which takes the system off the manifold, and projection back to the manifold by optimizing a local cost function. To achieve this we maximize the agreement of local subsystems over a finite length (patch size) $L$. We choose the manifold of uniform matrix product states (MPS) as our variational set, but instead of optimizing for global state fidelity as in established MPS methods such as time-evolving block decimation (TEBD)~\cite{vidalTEBD} or the time-dependent variational principle (TDVP)~\cite{haegemanTDVP}, we optimize the local subsystem agreement. This leads to a different trajectory on the manifold. We refer to this MPS-based algorithm as locally optimized matrix-product states (LOMPS); the implementation used in this work is available in Ref.~\cite{lomps_code}.

Our approach is part of a growing effort to describe quantum dynamics through the information that remains locally accessible. Locality, hydrodynamic relaxation, effective dissipation, and operator spreading all provide mechanisms by which local dynamics can become simpler than the evolving global wave function
\cite{BravyiHastingsVerstraete,OperatorHydrodynamics,NahumOperatorSpreading,KhemaniDissipativeHydrodynamics,RakovszkyDiffusiveOTOC,XuSwingleScrambling,ParkerOperatorGrowth}; see also Refs.~\cite{DAlessioETH,RandomQuantumCircuitsReview}. Some examples particularly closely related include works on density-matrix truncation \cite{DensityMatrixTruncation}, dissipation-assisted operator entanglement (DAOE) and operator backflow \cite{DAOE,OperatorBackflow}, entanglement-to-mixture conversion \cite{SuraceTrading,Converting}, MPS thermofield methods \cite{MPSThermofield}, information-lattice and local-RDM hierarchy methods \cite{InformationLattice,InfoLattice,CoMPuTE}, semiclassical phase-space and truncated-Wigner approaches \cite{PolkovnikovPhaseSpace,SchachenmayerDTWA,WurtzCTWA,WurtzQuantumDiffusion}, transverse light-cone contractions \cite{LightConeTN}, influence-matrix approaches \cite{InfluenceMatrix}, and tensor-network calculations based on artificial dissipation \cite{StoudenmireJC}.

Within this broader landscape, our approach is distinguished by the combination of three features. Firstly, the classical algorithm uses computational resources efficiently by trading global wave-function fidelity for an accurate description of local observables. Secondly, the classical algorithm can be turned into a quantum-classical hybrid algorithm that is robust to hardware noise and allows exploration of a potential intensive quantum advantage. Finally, the variational formulation provides both an analytic understanding and a geometric perspective on the dynamics, including closed-form equations of motion.

Our results demonstrate that the local variational algorithm reproduces coherent early-time dynamics, captures late-time thermalisation, and show proof-of-principle implementations on both IBM hardware and Quantinuum H2. Looking ahead, the local formulation may provide a direct route to identifying thermal fixed points and preparing locally thermal states \cite{Temme2011QuantumMetropolis,Motta2020QITE,Ding2025KMSSampler,Rouze2026OptimalGibbs,Ding2026ThermalPreparation} efficiently. The algorithm opens up a route towards a geometric understanding of local thermalisation and intrinsic complexity of many-body dynamics. The quantum-classical hybrid algorithm could be particularly valuable in two dimensions, where the reduced density matrices required by a classical implementation rapidly become too large to handle.

\section{Results}

We demonstrate the three core features of our proposal, early-time coherence, late-time thermalization and a proof-of-principle demonstration of the quantum algorithm on a typical quantum quench with an integrable and a non-integrable spin chain, the quantum Ising model 
\begin{equation}
  H_{\mathrm{Ising}}
  =
  -\sum_i
  \left(
  \sigma_i^z\sigma_{i+1}^z
  +g\sigma_i^x + h \sigma_i^z
  \right)\;,
  \label{eq:tfim}
\end{equation}
which is integrable in the transverse-field limit ($h=0$) and non-integrable otherwise. Depending on the energy density of the initial state, the system is known to show weak or strong thermalization \cite{banuls2011strong}.

The variational algorithm we introduce is based on locally optimizing a matrix product state ansatz and referred to as LOMPS. We use a left-canonical uMPS ansatz and update the MPS tensor by optimizing the Hilbert-Schmidt distance of the $L$-site reduced density matrix after each Trotter time step.  
To accurately capture the local reduced density matrices we work with tensors of large enough bond dimension $D\sim 2^L$ (see App.~\ref{app:lomps-representability}). We find that the $L$-site algorithm accurately reproduces $l$-site observables for $l<L$ and increasing the patch size $L$ systematically improves their accuracy. 
\subsection{Early-time coherence}

To demonstrate early-time coherence, we consider a quantum quench in the
integrable transverse-field Ising model from the paramagnetic ground state at $g_i=1.5$ to the
ferromagnetic regime $g_f=0.2$.  We diagnose the coherent local response
using the local Loschmidt signature
\begin{equation}
  \lambda_L(t)
  =
  -\frac{1}{L}
  \log
  \Tr\!\left[
  \rho_L(t)\rho_L(0)
  \right],
  \label{eq:local-loschmidt}
\end{equation}
where \(\rho_L(t)\) is the \(L\)-site reduced density matrix and compare it to an exact result obtained via mapping to free fermions~\cite{PeschelRDM}\footnote{The free fermion result is quasi exact and obtained via a finite, but large system size $N=128$ calculation.} (cf. Fig.~\ref{fig:main_figure}(c)). We observe that the local Loschmidt echo
retains the characteristic signature of the dynamical quantum phase transition for early times.

\subsection{Late-time thermalization}

\begin{figure}[H]
\centering
\includegraphics[width=\columnwidth]{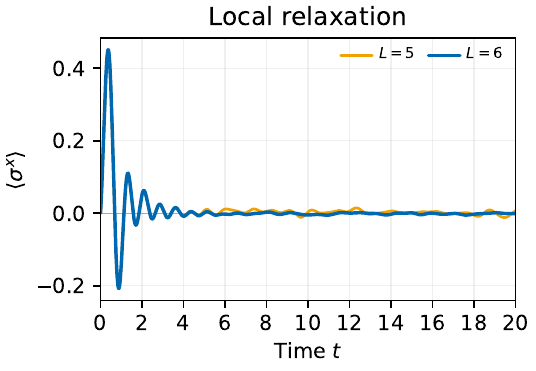}
\caption{\textbf{Thermalization of a local observable.} Single-site
\(\langle\sigma^x\rangle\) dynamics in the tilted-field Ising quench. Increasing the patch sizes reduces artificial late-time oscillations.}
\label{fig:relax}
\end{figure}

To study thermalising dynamics, we consider a strongly thermalizing quench in a non-integrable system: the tilted-field Ising chain with \(g=-1.05\) and
\(h=0.5\). 
We start from a $y$-polarized
product state, that has zero energy density and is known to relax to the infinite temperature thermal state~\cite{banuls2011strong}. The LOMPS algorithm performed for a patch size $L$ produces trajectories for which the interior $l$-site sub systems ($l<L$) show strong convergence to the expected thermal steady state, as signified by their entanglement entropy reaching maximal values (cf. Fig.~\ref{fig:main_figure}(d)). The algorithm captures the relaxation of local observable more accurately, when we increase $L$. Figure~\ref{fig:relax} demonstrates that increasing the patch size from $L=5$ to $L=6$ suppressed residual late-time oscillations of $\langle\sigma^x\rangle$ and improves the late-time agreement with its expected infinite-temperature value. For this quench, LOMPS also retains an accurate
description of the local dynamics for longer than iTEBD at comparable
representation memory. Systematic comparisons with iTEBD and results for
varying patch size are presented in
App.~\ref{app:systematic-improvement}.

\subsection{Quantum-Classical Algorithm}

\begin{figure*}[!t]
  \centering
  \includegraphics[width=\textwidth]{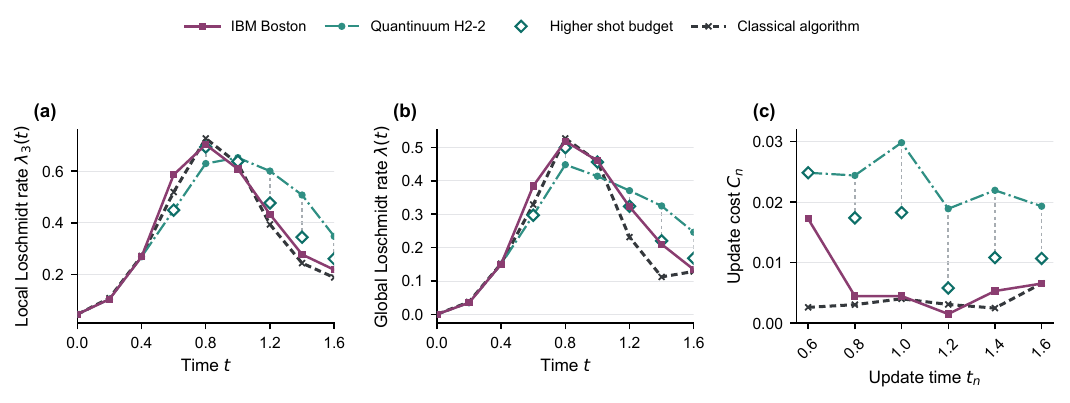}
  \caption{\textbf{Quantum algorithm.}
  \textbf{(a)} Three-site local Loschmidt rate,
  \textbf{(b)} global Loschmidt rate and
  \textbf{(c)} cost function
obtained by running the variational algorithm on IBM Boston Heron r3 with 5000 shots and on Quantinuum H2-2 with 500 shots per circuit evaluation, compared to the noiseless classical
  algorithm with the same MPS bond dimension. The first three states, at
  $t=0,0.2,0.4$, were supplied by a warm start. Diamonds show the expected improvement of the
  Quantinuum optimization protocol performed with a higher shot budget (for details see App.~\ref{sec:diamond}).}
  \label{fig:hardware-trajectories}
\end{figure*}

The quantum-classical counterpart of our classical variational algorithm evaluates the cost function
on a quantum computer and performs the optimization on a classical computer. To measure the cost function we use the sequential circuit preparation of MPS with a resource efficient parametrization of bond dimension $D=2$ MPS tensors and perform swap measurements to compute overlaps of reduced density matrices that define the local cost function as detailed in the next section (cf. Eq.~\eqref{eq:qdmt-cost}). We executed the algorithm for the transverse-field Ising model quench on both Quantinuum's H2-2 processor and IBM's Heron r3 processor and computed the local, cf. Eq.~\eqref{eq:local-loschmidt}, and global, $\lambda(t)=- \lim_{N \to \infty} \frac{1}{N} \log |\braket{\Psi(t)|\Psi(0)}|^2$, Loschmidt signatures obtained from the recorded parameter trajectory (see Fig.~\ref{fig:hardware-trajectories}). On both devices the algorithm clearly recovers the qualitative features.

\section{Algorithm}
\label{sec:algorithm}
We now 
define the classical algorithm, present its
continuous-time equations of motion and its quantum counterpart.

	\subsection{Classical algorithm}
\label{sec:classical-algorithm}

The classical algorithm uses a uniform matrix product state (MPS) $\ket{\Psi(A(t))}$ as its underlying ansatz (see App.~\ref{app:lomps} for more technical details). The variational time evolution algorithm consists of a two-step protocol. First, we time-evolve the state $\ket{\Psi(A)}$ by a Trotter time step $\Delta t$. Second, we project back to the variational manifold by minimizing the Hilbert-Schmidt distance of the $L$-site reduced density matrix of the time-evolved state and the ansatz state $\ket{\Psi(A')}$.

\subsubsection{Time evolution}

\begin{figure}[H]
\centering
\includegraphics[width=\columnwidth]{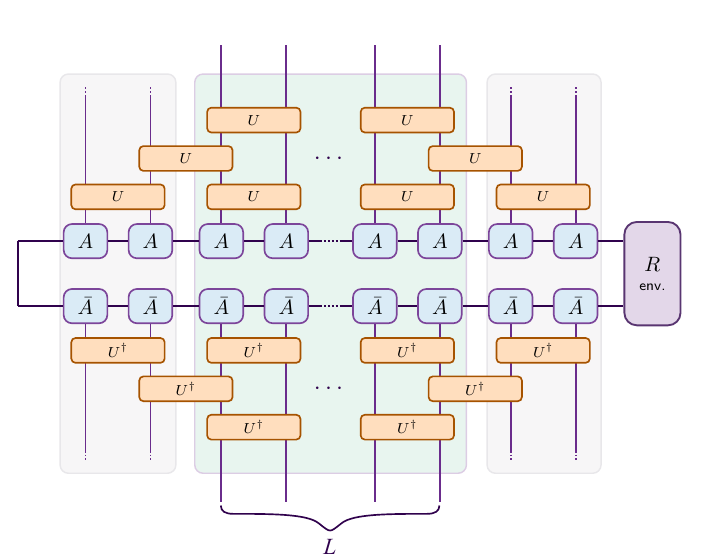}
\caption{\textbf{Time evolution.}  The classical algorithm time evolves the state $\ket{\Psi(A)}$ by a short Trotter circuit and constructs its local reduced density matrix on $L$ sites $\rho_L^\text{evol}(t+\Delta t)$ (green) by tracing out the environment (purple).}
\label{fig:cost-equality}
\end{figure}

Starting from \(\ket{\Psi(A(t))}\), we time evolve by a $k$-th order Trotter step unitary $U({\Delta t})$ \cite{trotter1959product,SUZUKI1993232}. As our variational algorithm is designed to optimize local agreement between states, we then construct the $L$-site reduced density matrix $\rho_L^{\mathrm{evol}}(t+\Delta t)$ of the time evolved state by tracing out an infinite left and right environment. As illustrated in Fig.~\ref{fig:cost-equality} this amounts to time evolving an enlarged (buffered) region with a finite Trotter light cone~\cite{LocalTrotterErrors}.

\subsubsection{Projection}
\begin{figure*}[t]
\centering
\includegraphics[width=0.85\textwidth]{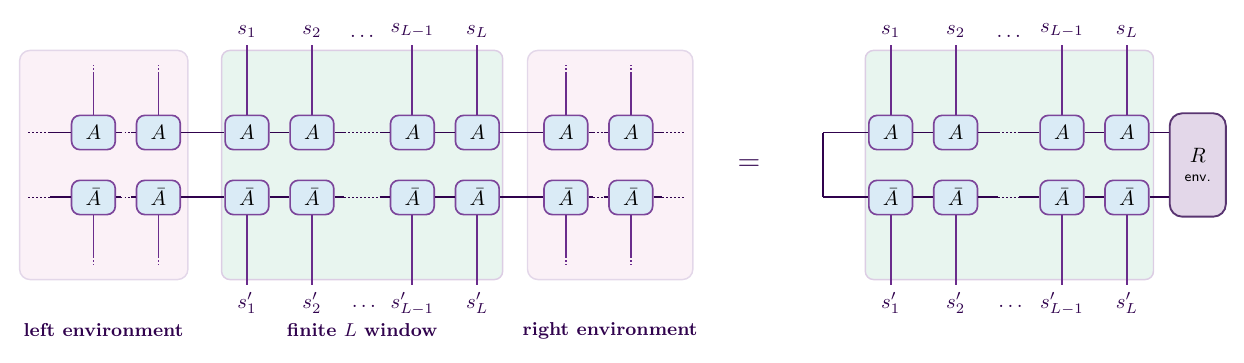}
\caption{\textbf{Local reduced density matrix of a uniform MPS.}  The left and
right environments of an infinite uniform (i.e., translation-invariant) MPS reduce to the left and right fixed point of the transfer matrix (see App.~\ref{app:lomps-umps}) for details. Using the left-canonical gauge, the left-fix point is the identity and the right fixed point $R_\text{env}$ is computed explicitly.}
\label{fig:ansatz-rdm-equality}
\end{figure*}

	\begin{equation}
  C_L(A')
  =
  \frac12
  \left\|
  \rho_L[A']-\rho_L^{\mathrm{evol}}(t+\Delta t)
  \right\|^2_\text{HS}
  \label{eq:qdmt-cost} \;,
\end{equation}
where $|| \cdot ||_\text{HS}$ is the Hilbert-Schmidt distance and the reduced density matrix $\rho[A']$ is obtained from the ansatz state $\ket{\Psi(A')}$ by tracing out a left- and right-environment as shown in Fig.~\ref{fig:ansatz-rdm-equality}.

\subsubsection{Cost function}
While other reduced density matrix distances could be used, our particular choice is motivated by three observations. Simplicity: the Hilbert-Schmidt distance is easy to compute numerically and leads to closed-form equations of motion for the variational parameters. Compatibility with quantum measurements: the Hilbert-Schmidt distance of reduced density matrices can be inferred via measurements of reduced density matrix overlaps, a feature that allows us to execute our variational algorithm on a quantum computer. An intuitive interpretation: The cost function splits into three terms  
$$C_L(A')=\frac 1 2 \left(P_\text{evol} - 2 F  + P' \right)\;,$$
the purity of the target state, $P_\text{evol}= \Tr \left(\rho_L^\text{evol}(t+\Delta t)^2\right)$, which is constant during a time step and need not be computed, the 
\begin{itemize}
	\item \emph{local fidelity} $F:=\Tr \left(  \rho_L^{\mathrm{evol}}\rho_L[A']
\right)$ and the
	\item \emph{ansatz purity} $P':=\Tr \left(  \rho_L[A']^2
\right)$ \;.
\end{itemize}
The local fidelity must be maximized while the ansatz purity must be minimized. As a consequence the emerging reduced density matrix $\rho(A')$ balances retaining local agreement with the target state against acquiring admixture which is a desired feature for a thermalization assisted variational algorithm. Note, that the length scale $L$ interpolates between a purely local, short-sighted optimization problem and established variational algorithms relying on global fidelity such as TDVP. Though not the focus of our discussion, this algorithm has a natural generalization to inhomogeneous systems.

The details of how to numerically implement the optimization are discussed in App.~\ref{app:lomps}. 

\subsection{Equations of motion}
\label{sec:eom}

Complementary to the discrete numerical algorithm, our protocol also implies
continuous-time equations of motion for the variational parameters. For real
local coordinates \(\lambda^i\), the continuous-time limit of the local
projection gives
\begin{equation}
  \Tr_L\!\left(\partial_i\rho_L\,\partial_j\rho_L\right)\dot\lambda^j=-i\,\Tr_L\!\left(\partial_i\rho_L\,\Tr_{\bar L}[H,\rho]\right).
  \label{eq:qdmt-eom}
\end{equation}
A derivation is given in App.~\ref{app:lomps-continuous-time}.

This equation is the local analogue of the TDVP equation of motion~\cite{haegemanTDVP}. TDVP uses the
global wavefunction metric while for LOMPS the relevant metric $
  S_{ij}
  =
 \Tr\!\left(
  \partial_i\rho_L\,\partial_j\rho_L
  \right)$ is induced by how the variational parameters $\lambda^i$ parametrize the local $L$-site reduced density matrices. 

\subsection{Quantum algorithm}
\label{sec:quantum-algorithm}

\begin{figure}[t]
\centering
\includegraphics[width=0.46\textwidth]{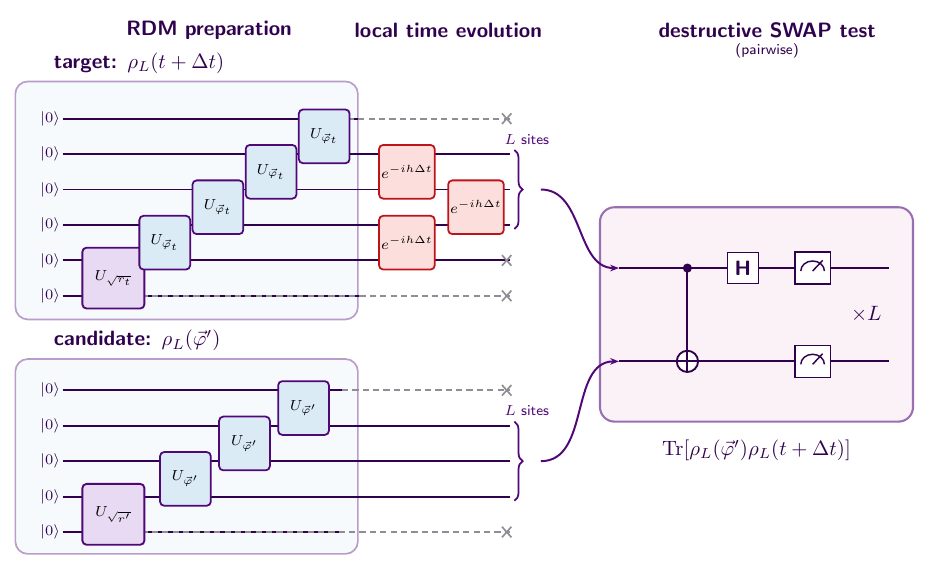}
\caption{\textbf{Evaluation of the local cost function on a quantum computer.}  We prepare (an open boundary version of) the state $\ket{\Psi(\vec \varphi_t)}$ and time evolve it. Its $L$-site interior together with the right-environment $\sqrt{r}$ defines the target reduced density matrix $\rho_L(t+\Delta t)$. Simultaneously, we prepare the candidate state $\ket{\Psi(\vec \varphi')}$ that defines the ansatz reduced density matrix $\rho_L(\vec \varphi')$. 
The overlap of $\rho_L(t+\Delta t)$ and $\rho_L(\vec \varphi')$ is measured by a destructive SWAP test. The evaluation of the ansatz purity proceeds via an analogous protocol.}
\label{fig:quantum-overlap-circuit}
\end{figure}

When evaluating the cost function on a quantum computer it is sufficient to focus on the local fidelity and the ansatz purity, both of which are given by overlaps of local reduced density matrices. To prepare and measure these we built upon the sequential circuit preparation of MPS
\cite{PhysRevLett.95.110503,PhysRevA.75.032311,barratt2021parallel,
dborinSimulatingGroundstateDynamical2022}. However, we do not prepare a finite-size MPS, but an open boundary version that is sufficient to construct the time-evolved state and trace out the left-and right environment as shown in Fig.~\ref{fig:quantum-overlap-circuit} and explained in more detail in App.~\ref{app:quantum-algorithm-methods}.
The circuit and optimization code used for the hardware demonstrations is
available in Ref.~\cite{qlomps_code}.

We focus on an MPS ansatz with bond dimension \(D=2\). After quotienting out
irrelevant gauge transformations and the common tensor phase, the full
physical manifold of qubit \(D=2\) uniform MPS has eight independent real
parameters. In the TFIM quench, we can further reduce the parametrization to six independent angles, $\vec \varphi$, and prepare the uniform MPS state with a 2-qubit unitary gate $U(\vec \varphi)$ composed of two CNOT gates and single-qubit rotations (cf. Fig.~\ref{fig:six-angle-circuit}). 

\emph{Local fidelity.} To evaluate the local fidelity for a time step $t\to t+\Delta t$ we prepare the current time state $\ket{\Psi(\vec \varphi_t)}$ and time evolve it by a Trotterized time evolution. Simultaneously, we prepare the ansatz state $\ket{\Psi(\vec \varphi'})$ on a different set of qubits. We then perform a destructive swap test~\cite{swap} on all $L$ pairs of the
central \(L\) qubits of the two states (see Fig.~\ref{fig:quantum-overlap-circuit}) to compute the overlap  $\tr (\rho_L(t+\Delta t) \rho_L(\vec \varphi'))$ of the reduced density matrices. 

\emph{Ansatz purity.} To measure the ansatz purity $\tr (\rho_L(\vec \varphi') \rho_L(\vec \varphi'))$ we proceed analogously. We prepare two copies of $\ket{\Psi(\vec \varphi')}$ and perform the destructive swap test on all $L$ pairs of the central qubits.

For each time step in the evolution we perform this protocol multiple times for different ansatz parameters. The results of the cost function measurements are fed into a classical algorithm that chooses a new candidate set. After a finite amount of iterations this results in a candidate for a local minimum $\vec \varphi_\text{min}$ of the cost function. This defines the variational state at the next time step $\vec \varphi_{t+\Delta t} := \vec \varphi_\text{min}$. The details on how to choose the candidates and how to infer the candidate for the local minimum are provided in App.~\ref{app:quantum-algorithm-methods}.

\section{Discussion}

We have introduced a variational approach to many-body dynamics that evolves the information required to describe local observables rather than attempting to preserve the global wave function. By matching reduced density matrices on finite regions, LOMPS reproduces coherent early-time dynamics and captures late-time thermalisation, with accuracy improving systematically as the optimised region is enlarged. The same variational principle admits a quantum-classical implementation, and our proof-of-principle experiments on IBM and Quantinuum hardware recover the characteristic local dynamics despite shot noise and device imperfections.

These results suggest an operational notion of intrinsic local complexity. For a specified observable, accuracy, and evolution time, this complexity can be associated with the spatial range and variational resources required for convergence --- for LOMPS, the patch size $L$ and the bond dimension $D$. In this view, increasing the patch size does not simply improve a global approximation: it increases the range over which dynamically relevant information is retained. The quantum implementation offers a complementary route when representing the corresponding local reduced density matrices becomes classically costly, although establishing the scaling with the patch size and the bond dimension beyond the present proof of principle will be necessary before a practical quantum advantage can be assessed.

The same interpretation identifies the principal limitation of finite-window evolution. Matching an $L$-site reduced density matrix at the current time does not uniquely determine the larger reduced density matrices that can affect future evolution, so a small local projection cost is not by itself a guarantee of trajectory accuracy. Understanding when this unresolved information remains irrelevant to the observables of interest is therefore central to determining the regime of validity of the method. Diagnostics of this regime could enable adaptive choices of patch size, bond dimension, or time step. The continuous-time equations of motion provide another route to this question: their geometry, fixed points, and slow modes may reveal how the relevant spatial and temporal scales change between strongly thermalising, weakly thermalising, and nonthermalising dynamics.

More broadly, our results support a shift in perspective on the simulation of many-body dynamics. The exponential complexity of an evolving global wave function need not directly determine the difficulty of predicting local observables. What matters instead may be the amount of information that must remain locally accessible over the relevant space and time scales. Thermalisation provides an especially favourable setting because this information can simplify even while the global state continues to become more complex. Local variational evolution provides both a computational strategy for exploiting this separation and a framework for asking how large this intrinsic local complexity actually is.

\begin{acknowledgments}
This work was supported by the EPSRC UK under
Grants No.~EP/Z53318X/1, EP/S005021/1, EP/S021582/1.
We acknowledge Quantinuum and the National Quantum Computing Centre (NQCC)
for providing access to quantum-computing resources used in this work.
\end{acknowledgments}

\clearpage

\appendix

\section{Further Results}

\subsection{Operator-space complexity of local reduced states}
\label{app:osee}

\begin{figure}[t]
  \centering
  \includegraphics[width=\columnwidth]{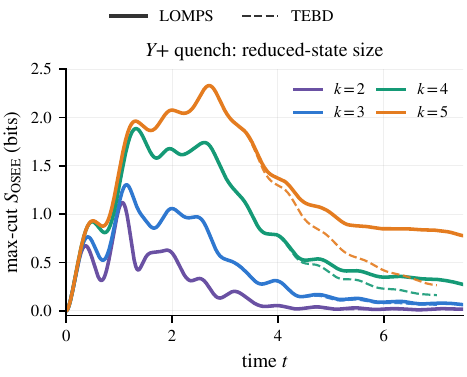}
  \caption{\textbf{Operator-space complexity of local reduced states.}
  Maximum-cut operator-space entanglement entropy (OSEE) for \(k=2,\ldots,5\)-site reductions of the
  \(L=5\) LOMPS trajectory for the strongly thermalizing \(Y+\) quench.
  Solid curves show LOMPS and dashed curves the available TEBD reference.}
  \label{fig:local-osee}
\end{figure}

To quantify the local description-complexity in the cartoon in
Fig.~\ref{fig:main_figure}(a), we use the operator-space entanglement entropy
(OSEE) of finite reduced density matrices
\cite{ProsenPizornOSEE}.  We vectorize a \(k\)-site reduced state
\(\rho_k\), normalize it in the Frobenius norm, and Schmidt-decompose the
result across each spatial cut.  The maximum-cut OSEE is obtained by
\begin{equation}
  \begin{aligned}
    S_{\mathrm{OSEE}}(\rho_k)
    &=
    \max_{1\leq j<k}
    \left[
      -\sum_\alpha p_{\alpha,j}\log_2 p_{\alpha,j}
    \right],
    \\
    p_{\alpha,j}
    &=
    \frac{s_{\alpha,j}^2}{\sum_\beta s_{\beta,j}^2},
  \end{aligned}
  \label{eq:osee-definition}
\end{equation}
where \(s_{\alpha,j}\) are the
operator Schmidt values across the cut. Figure~\ref{fig:local-osee} shows the OSEE for the strongly thermalizing \(Y+\) quench obtained via LOMPS and TEBD for various local reduced density matrices and shows qualitative agreement with the heuristic picture.

\subsection{Systematic improvement with local resolution}
\label{app:systematic-improvement}

\begin{figure*}
  \centering
  \includegraphics[width=\textwidth]{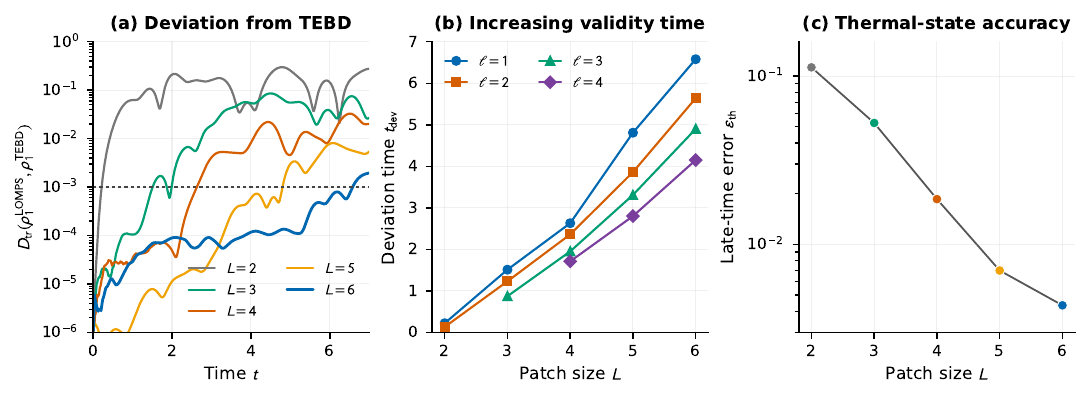}
  \caption{\textbf{Systematic improvement with the fitted patch size.}
  Results for the non-integrable Ising quench from the \(y\)-polarized product
  state, using the same time step and accepted projection-cost threshold for
  \(L=2,\ldots,6\).
  \textbf{(a)} Trace distance between the one-site LOMPS and TEBD reduced
  density matrices.  The dotted line marks \(10^{-3}\).
  \textbf{(b)} First threshold-crossing time \(t_{\mathrm{dev}}\) for
  \(\ell=1,\ldots,4\)-site reduced density matrices with \(\ell\leq L\).  
  \textbf{(c)} Late time deviation (cf.~Eq.~\eqref{eq:late-time-thermal-error}) of the one-site LOMPS reduced density matrix from
  the infinite-temperature state \(I/2\).}
  \label{fig:yplus-systematic-improvement}
\end{figure*}

We quantify the convergence of LOMPS with the fitted patch size using the
non-integrable Ising quench from the \(y\)-polarized product state introduced
in the main text.  We compare trajectories with \(L=2,\ldots,6\), all evolved
with time step \(\Delta t=10^{-3}\) and the same accepted per-step projection
cost \(C\leq10^{-12}\). The bond dimensions are chosen such that the number of free parameters of the left-canonical tensor $A$ exceeds that of the expected $L$-site reduced density matrices (cf. Eq.~\eqref{eq:lomps-draft-representability-count}) by a significant margin (\(D=5,7,12,21,41\)).  We compare the LOMPS data to a 
quasi-exact iTEBD reference calculation using a growing bond dimension up to
\(t_{\mathrm{TEBD}}\simeq 7\), where the reference reaches \(\chi=1628\). At
later times, for which this reference is unavailable, we instead examine how
the LOMPS trajectories change as \(L\) is increased.

For the direct comparison with iTEBD, we use the trace distance
\begin{equation}
  \mathcal D_\ell(t;L)
  =
  \frac{1}{2}
  \left\|
    \rho_\ell^{\mathrm{LOMPS}}(t;L)
    -\rho_\ell^{\mathrm{TEBD}}(t)
  \right\|_1 .
  \label{eq:rdm-trace-distance}
\end{equation}
Figure~\ref{fig:yplus-systematic-improvement}(a) shows that the one-site reduced density matrix
remains close to the iTEBD result for progressively longer times as \(L\) is
increased.  We summarize this behavior by defining the deviation time
\(t_{\mathrm{dev}}(\ell;L)\) as the first sampled time at which
\begin{equation}
  \mathcal D_\ell(t;L)>10^{-3}.
  \label{eq:rdm-threshold-time}
\end{equation}
At every length \(\ell\) of the reduced density matrix, increasing the LOMPS patch size \(L\) delays this threshold
crossing, as shown in Fig.~\ref{fig:yplus-systematic-improvement}(b). The scaling of the delay vs $L$ depends on the chosen threshold value, but is approximately linear for a wide range of threshold values. 

\begin{figure*}
  \centering
  \includegraphics[width=0.82\textwidth]{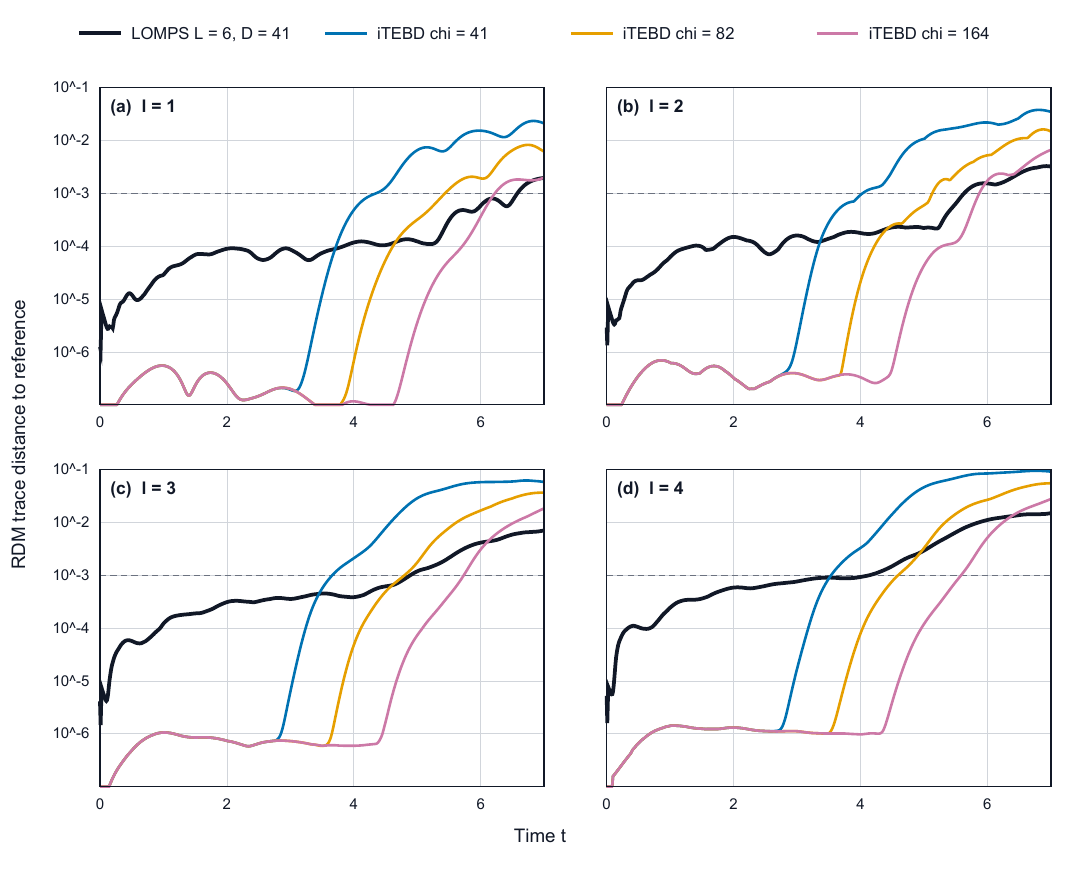}
  \caption{\textbf{Comparison with finite bond-dimension iTEBD.}
  Trace distance of the \(\ell\)-site reduced density matrix from the
  growing-bond-dimension iTEBD reference for the non-integrable Ising quench
  from the \(y\)-polarized product state.  We compare LOMPS with \(L=6\) and
  \(D=41\) to iTEBD calculations capped at \(\chi=41,82\), and \(164\).
  The panels show \(\ell=1,\ldots,4\), and the dotted line marks
  \(10^{-3}\).}
  \label{fig:capped-itebd-comparison}
\end{figure*}

The growing-bond-dimension TEBD calculation used above provides a
quasi-exact reference, but does not represent a simulation with a fixed
computational budget.  We therefore also compare the \(L=6\), \(D=41\)
LOMPS trajectory with iTEBD calculations whose bond dimensions are limited to
\(\chi=41,82\), and \(164\).  Figure~\ref{fig:capped-itebd-comparison} shows
the trace distance of each trajectory from the same growing-bond-dimension
iTEBD reference.  At the end of the accessible reference interval,
\(t=7\), LOMPS has the smaller error for every reduced-state size shown,
\(\ell=1,\ldots,4\), including in comparison with the \(\chi=164\) iTEBD
calculation.  This comparison concerns the accuracy obtained at fixed bond
dimension.  Since the memory required to represent an MPS is controlled by
its bond dimension, it allows a meaningful comparison of representation
memory, but not a direct comparison of computational runtime.

We separately test the late-time behavior using
the independently known thermal state.  The quench has zero energy density
and is expected to relax locally to infinite temperature state
\(\rho_1^{\mathrm{th}}=I/2\).  We define the late-time error
\begin{equation}
  \begin{aligned}
    \epsilon_{\mathrm{th}}(L)
    &=
    \Bigg[
      \frac{1}{t_2-t_1}
      \int_{t_1}^{t_2}\!\dd t
      \\
    &\qquad
      \mathcal D_{\mathrm{tr}}\!\left(
        \rho_1^{\mathrm{LOMPS}}(t;L),\frac{I}{2}
      \right)^2
    \Bigg]^{1/2},
  \end{aligned}
  \label{eq:late-time-thermal-error}
\end{equation}
and choose \(t_1=10\) and \(t_2=20\). The late-time error decreases exponentially with $L$ over the range considered as shown in Fig.~\ref{fig:yplus-systematic-improvement}(c). Thus increasing 
the patch size both extends agreement with the reference dynamics and
improves the eventual approach to the expected thermal state.

\FloatBarrier
\subsection{Dependence on the thermalization regime}
\label{app:thermalization-regime}

\begin{figure*}[p]
  \centering
  \includegraphics[width=0.98\textwidth]{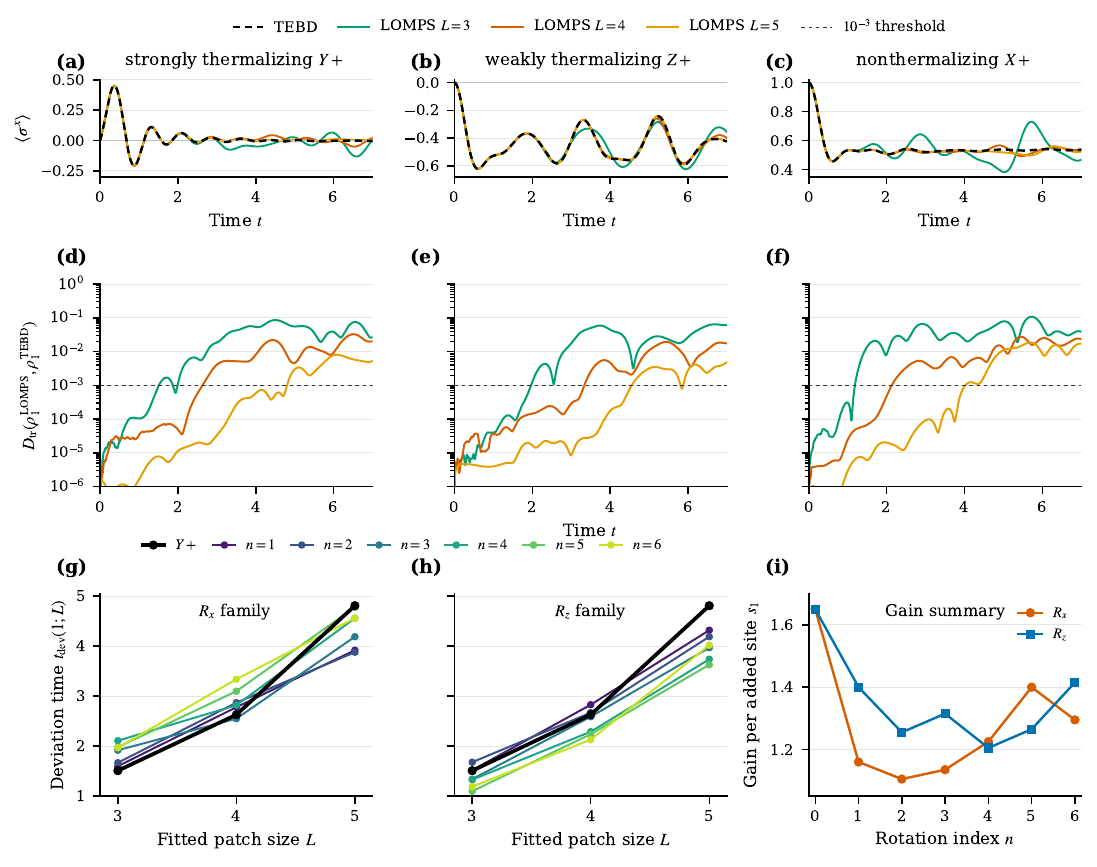}
  \caption{\textbf{Systematic improvement across thermalization regimes.}
  \textbf{(a)--(c)} Local observable $\langle\sigma^x\rangle$ for the
  strongly thermalizing $Y+$, weakly thermalizing $Z+$, and nonthermalizing
  $X+$ quenches, respectively.  Dashed curves are TEBD and solid curves are
  LOMPS.  \textbf{(d)--(f)} Corresponding one-site trace distances
  $\mathcal D_1(t;L)$; the dotted line marks the $10^{-3}$ threshold.
  \textbf{(g),(h)} Deviation time $t_{\mathrm{dev}}(1;L)$ for $Y+$ and all
  $n=1,\ldots,6$ members of the $R_x$ and $R_z$ families.  Its increase with
  $L$ demonstrates systematic improvement throughout the scan.
  \textbf{(i)} Gain $s_1$ from Eq.~\eqref{eq:patch-gain-rate}.}
  \label{fig:banuls-thermalization-scan}
\end{figure*}

We test whether the systematic improvement with $L$ extends beyond the
strongly thermalizing $Y+$ quench using two representative initial-state
families introduced in Ref.~\cite{banuls2011strong}, which rotate $Y+$ toward
the weakly thermalizing $Z+$ and nonthermalizing $X+$ quenches, respectively:
\begin{equation}
  \begin{aligned}
    \lvert\psi_n^{(x)}\rangle
    &=R_x(n\pi/12)\lvert Y+\rangle,\\
    \lvert\psi_n^{(z)}\rangle
    &=R_z(-n\pi/12)\lvert Y+\rangle,
    \qquad n=1,\ldots,6,
  \end{aligned}
  \label{eq:banuls-quench-scans}
\end{equation}
where $R_\alpha(\theta)=\exp(-i\theta\sigma^\alpha/2)$.  The LOMPS
trajectories use $(L,D)=(3,7),(4,12),(5,21)$, \(\Delta t=10^{-3}\), and the
same accepted projection-cost threshold $C\leq10^{-12}$.  Applying the observable and
reduced-density-matrix diagnostics established above, Fig.~\ref{fig:banuls-thermalization-scan}
shows that increasing $L$ systematically extends agreement with TEBD for
every quench tested.

Using Eq.~\eqref{eq:rdm-threshold-time}, we define the gain obtained by
increasing $L$ from $3$ to $5$, per added fitted site, as
\begin{equation}
  s_1=
  \frac{t_{\mathrm{dev}}(1;5)-t_{\mathrm{dev}}(1;3)}{2}.
  \label{eq:patch-gain-rate}
\end{equation}
The strongly thermalizing $Y+$ quench gives $s_1=1.65$, compared with
$1.10$--$1.42$ across the rotated states.  This suggests a mild tendency for
larger gains in more strongly thermalizing regimes, but not a universal
dependence.

\clearpage

\section{Quantum implementation details}
\label{app:quantum-algorithm-methods}

\subsection{Sequential-circuit MPS ansatz.}
In the quantum-classical implementation of our local variational algorithm we prepare the state which defines the reduced density matrix $\rho_L(\vec \varphi)$ by a sequential MPS preparation \cite{PhysRevLett.95.110503,PhysRevA.75.032311,barratt2021parallel,
dborinSimulatingGroundstateDynamical2022}. This proceeds via the identification of the MPS tensor $A$ and the unitary $U$
\begin{equation}
  A^s_{\alpha\beta}(\vec \varphi)
  =\langle\alpha,s\rvert U(\vec \varphi)
  \lvert0,\beta\rangle,
  \label{eq:sequential-isometry}
\end{equation}
where \(\beta\) is the incoming bond index, \(s\) the emitted physical index and
\(\alpha\) the outgoing bond index.  Unitarity of \(U\) makes the tensor left
canonical by construction.  At time step \(n\), the parameters
\(\vec \varphi_n\) prepare the current MPS.  A fixed short-time circuit
generates the buffered target, while otherwise identical circuits prepare
candidate states at parameters \(\vec \varphi\).

\begin{figure}[b]
  \centering
  \includegraphics[width=\columnwidth]{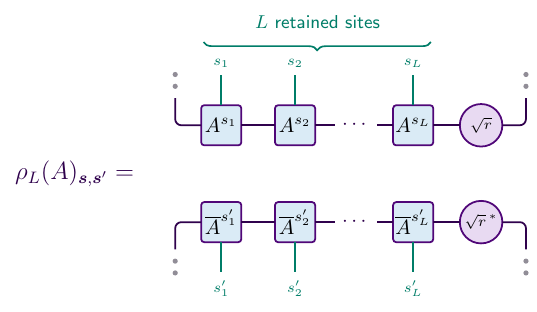}
  \caption{\textbf{Construction of the local reduced density matrix.}
  For a left-canonical uniform MPS, the semi-infinite environments reduce to
  the identity on the left and the normalized right fixed point \(r\).
  Writing \(r=\sqrt{r}\,\sqrt{r}^{\dagger}\), the right environment is
  represented by the factors \(\sqrt{r}\) and \(\sqrt{r}^{*}\) in the ket
  and bra layers. Contracting the virtual boundary indices while leaving the
  physical indices \(\boldsymbol{s}\) and \(\boldsymbol{s}'\) of the retained
  \(L\) sites open yields \(\rho_L(A)_{\boldsymbol{s},\boldsymbol{s}'}\).}
  \label{fig:RDM-r}
\end{figure}

The unitary is enough to specify the infinite uMPS, however, we need to represent reduced density matrices obtained after tracing out an infinite left and right environment. We implement this, by preparing a state supplemented with a left and right ancilla that, after tracing out the ancilla, results in the correct reduced density matrix. The construction is illustrated in Fig.~\ref{fig:RDM-r}. To prepare this state so, we need to compute the normalized right fixed-point of the transfer matrix $r$, with $\tr r =1$ and its square root $\sqrt{r}$. This factorization is always possible \cite{mps_representations}. In our protocol we compute $r$ and $\sqrt{r}$ on a classical computer and apply the unitary $U_{\sqrt{r}}$
\begin{equation}
  (\sqrt{r})_{\alpha\beta}(\vec \varphi)
  =\langle\alpha,\beta\rvert U_{\sqrt{r}}(\vec \varphi)
  \lvert0,0\rangle
  \label{eq:sequential-isometry-sqrt-r}
\end{equation}
to the two right-most qubits (see Fig.~\ref{fig:quantum-overlap-circuit}). The parametrizations of $U$ and $U_{\sqrt{r}}$ are now discussed.

\begin{figure*}[t]
\centering
\includegraphics[width=0.96\textwidth]{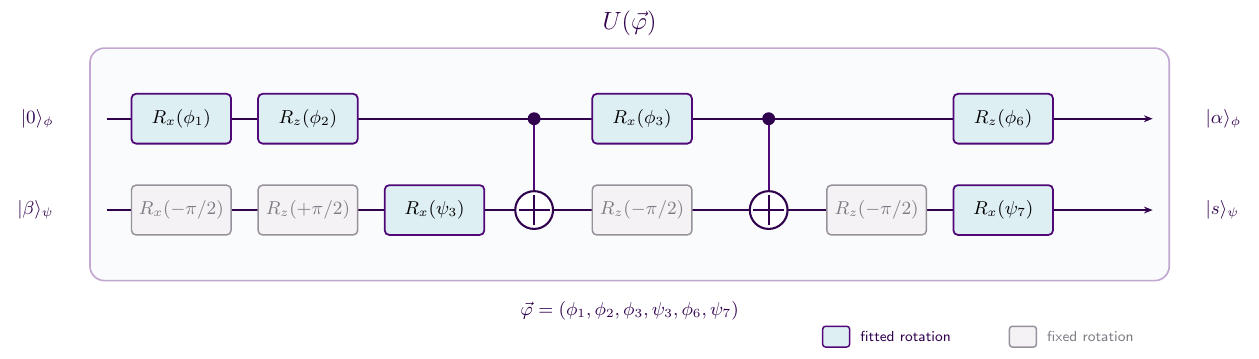}
\caption{\textbf{Six-angle uniform-MPS tensor unitary.}  The two-qubit circuit
implements \(U(\vec \varphi)\) in Eq.~\eqref{eq:six-angle-gate}, with gate
order read from left to right.  Blue gates contain the six fitted coordinates
\(\vec \varphi=(\phi_1,\phi_2,\phi_3,\psi_3,\phi_6,\psi_7)\); grey gates
are fixed basis rotations.}
\label{fig:six-angle-circuit}
\end{figure*}

\paragraph{Six-angle tensor unitary.}
For the transverse-field Ising quench, we found that a 6-parameter ansatz for the tensor $A$ was sufficient to represent the tensors along the early time trajectory ($t\leq 2$) to high accuracy as verified self-consistently by achieving low values of the cost function ($C_3 \leq 10^{-3}$) in the classical algorithm. The six parameters 
\begin{equation}
  \vec\varphi=(\phi_1,\phi_2,\phi_3,\psi_3,\phi_6,\psi_7)
  \label{eq:six-angles}
\end{equation}
define the unitary
\begin{align}
 U(\vec \varphi)={}&
 [R_z^{\phi}(\phi_6)\otimes R_x^{\psi}(\psi_7)]
 [\mathbb{I}\otimes R_z^{\psi}(-\pi/2)]
 \mathrm{CX}_{\phi\rightarrow\psi}
 \nonumber\\[-0.2em]
 &\times[R_x^{\phi}(\phi_3)\otimes R_z^{\psi}(-\pi/2)]
 \mathrm{CX}_{\phi\rightarrow\psi}
 [\mathbb{I}\otimes R_x^{\psi}(\psi_3)]
 \nonumber\\[-0.2em]
 &\times[R_z^{\phi}(\phi_2)\otimes R_z^{\psi}(\pi/2)]
 [R_x^{\phi}(\phi_1)\otimes R_x^{\psi}(-\pi/2)],
 \label{eq:six-angle-gate}
\end{align}
as shown in Fig.~\ref{fig:six-angle-circuit}. Equivalently, this is a thirteen-angle Euler
construction with
\(\psi_1=-\pi/2\), \(\psi_2=\pi/2\),
\(\psi_4=\psi_6=-\pi/2\), \(\psi_5=0\) and
\(\phi_4=\phi_5=0\). The restriction leaves six continuously varied angles and
two CNOT gates.  It was chosen to reduce the number of hardware-evaluated
directions in the present \(D=2\) proof of principle and does not imply that an
equally compact parametrization exists at larger bond dimension.

\paragraph{Right fixed-point boundary.}

For the right environment, we determine the normalized right fixed point
\(r(\vec\varphi)\),
\begin{equation}
  r(\vec\varphi)
  =\sum_s A^s(\vec\varphi)\,
  r(\vec\varphi)\,
  A^{s\dagger}(\vec\varphi),
  \qquad
  \Tr r(\vec\varphi)=1.
  \label{eq:quantum-right-fixed-point}
\end{equation}
For each parameter set, \(r(\vec\varphi)\) and its positive square root
\(\sqrt{r}(\vec\varphi)\) are computed classically. The unitary
\(U_{\sqrt r}(\vec\varphi)\) then prepares the two-qubit purification
\begin{equation}
  \lvert\Gamma_r(\vec\varphi)\rangle
  =
  U_{\sqrt r}(\vec\varphi)\lvert0,0\rangle
  =
  \sum_{\alpha,\beta}
  \bigl(\sqrt r\bigr)_{\alpha\beta}(\vec\varphi)
  \lvert\alpha,\beta\rangle.
  \label{eq:fixed-point-purification}
\end{equation}
Tracing out the reference qubit gives
\(\Tr_{\mathrm{ref}}\lvert\Gamma_r\rangle\langle\Gamma_r\rvert=r\), so this
boundary construction reproduces the right environment of the infinite MPS
without preparing a long chain. For \(D=2\), we diagonalize
\(r=V\operatorname{diag}(\lambda_0,\lambda_1)V^\dagger\), such that
\begin{equation}
  \lvert\Gamma_r\rangle
  =
  (V\otimes V^*)
  \left(
    \sqrt{\lambda_0}\lvert00\rangle
    +
    \sqrt{\lambda_1}\lvert11\rangle
  \right).
\end{equation}
The Schmidt coefficients are prepared by an \(R_y\) rotation through
\(2\operatorname{atan2}(\sqrt{\lambda_1},\sqrt{\lambda_0})\), followed by a
CNOT, while \(V\) and \(V^*\) are compiled into single-qubit Euler rotations
on the bond and reference qubits, respectively.

\paragraph{Product-layer target and translation average.}
To keep the quantum circuit shallow, we approximate the short-time evolution
by a single layer of two-site gates acting on one set \(\mathcal B\) of
disjoint nearest-neighbour bonds,
\begin{equation}
  U_{\mathrm{prod}}(\Delta t)
  =
  \prod_{j\in\mathcal B}
  e^{-ih_{j,j+1}\Delta t}.
  \label{eq:quantum-product-layer}
\end{equation}
Because the gates act on alternating bonds, this circuit distinguishes even
and odd sites and therefore has a two-site unit cell. An \(L\)-site window has
two inequivalent alignments relative to the gate layer, related by a one-site
translation. We denote the corresponding reduced density matrices by
\(\rho_{L,\mathrm L}^{\mathrm{prod}}\) and
\(\rho_{L,\mathrm R}^{\mathrm{prod}}\), and define the target reduced density matrix as their
average,
\begin{equation}
  \rho_L(t+\Delta t)
  =
  \frac{1}{2}
  \left(
    \rho_{L,\mathrm L}^{\mathrm{prod}}
    +
    \rho_{L,\mathrm R}^{\mathrm{prod}}
  \right).
  \label{eq:quantum-averaged-target}
\end{equation}
The two reduced density matrices are obtained from different placements of the observation window
within the same evolved circuit, rather than from two separate time
evolutions. Their average restores one-site translation invariance and defines
the target used in the quantum protocol.

\subsection{Measurement of the cost function}
For the averaged target reduced density matrix \(\rho_L(t+\Delta t)\) and candidate
\(\rho_L(\vec \varphi)\), the projection cost is one half of their squared
Hilbert--Schmidt distance,
\begin{align}
  2C_L(\vec \varphi)={}&
  \Tr[\rho_L(t+\Delta t)^2] \notag\\
  &+\Tr[\rho_L(\vec \varphi)^2] \notag\\
  &
  -2\Tr[\rho_L(t+\Delta t)\rho_L(\vec \varphi)].
  \label{eq:quantum-cost}
\end{align}
Writing
\(F_{\mathrm L,\vec \varphi_t}=\Tr[\rho_{L,\mathrm L}^{\mathrm{prod}}
\rho_L(\vec \varphi)]\) and
\(F_{\mathrm R,\vec \varphi_t}=\Tr[\rho_{L,\mathrm R}^{\mathrm{prod}}
\rho_L(\vec \varphi)]\), the part of the cost function that we need to minimize and measure is 
\begin{equation}
C_{\vec \varphi_t}(\vec \varphi)=P(\vec \varphi)-F_{\mathrm L,\vec \varphi_t}(\vec \varphi)-F_{\mathrm R,\vec \varphi_t}(\vec \varphi)  \;, \label{eq:eff-cost}
\end{equation} 
where
\(P=\Tr[\rho_L(\vec \varphi)^2]\).
Candidate purity and both target--candidate overlaps follow from
two-copy measurements of
\begin{equation}
  \Tr(\rho\sigma)=\Tr[\mathsf{S}(\rho\otimes\sigma)],
  \label{eq:swap-identity}
\end{equation}
where \(\mathsf{S}\) swaps the fitted registers.  We use the ancilla-free
destructive SWAP test of Ref.~\cite{swap}: corresponding
qubit pairs are rotated into the Bell basis by a CNOT and a Hadamard gate and
then measured.  The parity estimator \((-1)^{N_{11}}\), with \(N_{11}\) the number
of antisymmetric pair outcomes, averages to the required overlap.  Thus no
controlled-SWAP ancilla is required.

\paragraph*{Hardware execution and processing.}
On Quantinuum H2-2, each cost value was evaluated using a single
\(14\)-qubit circuit containing both the candidate-purity and
target--candidate-overlap measurements. The circuits were compiled to the
H2-2 gate set using the \textsc{pytket} Quantinuum backend. Intermediate
destructive-SWAP measurements were followed by qubit reset and reuse, as
described below. On IBM Boston, \(P\), \(F_{\mathrm L}\), and
\(F_{\mathrm R}\) were evaluated using three separate \(12\)-qubit circuits.
Each circuit was transpiled at optimization level \(2\) using the Qiskit
preset pass manager with a fixed physical-qubit layout and executed using
the Qiskit Runtime sampler.

For both devices, the overlaps were estimated directly from the measured
destructive-SWAP parities and combined according to
Eq.~\eqref{eq:eff-cost}. We applied no readout-error mitigation, zero-noise
extrapolation, offset subtraction, or shot-level postselection. The
acquisition-quality test used in the IBM protocol, described below, could
reject a complete scan affected by temporal drift, but it did not modify the
measurement outcomes retained from accepted scans.

\subsection{Shot-noise-aware variational update.}

\begin{figure*}[t]
\centering
\includegraphics[width=0.98\textwidth]{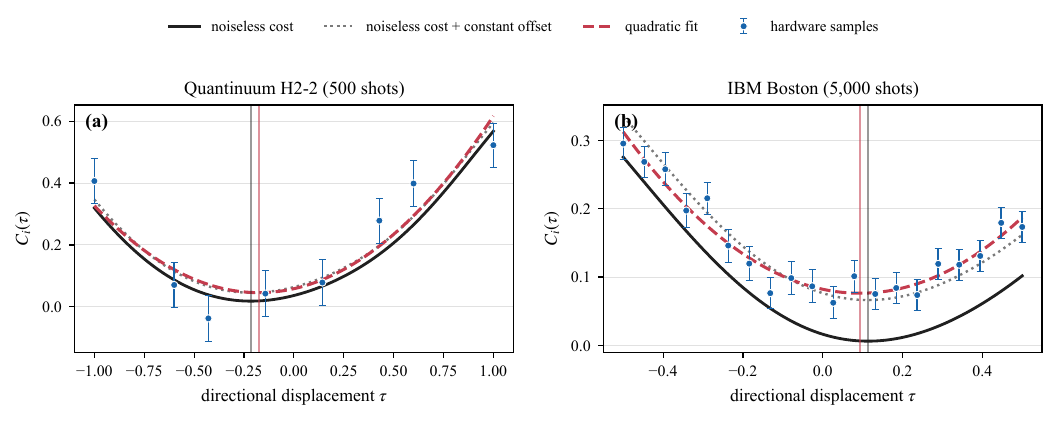}
\caption{\textbf{Quadratic parameter-cut samples on quantum hardware.}
Measured costs (blue points) along representative parameter cuts on
\textbf{(a)} Quantinuum H2-2, using 500 shots per cost evaluation, and
\textbf{(b)} IBM Boston, using 5,000 shots for each constituent circuit.
Dashed red curves show fits to Eq.~\eqref{eq:line-parabola}, solid black curves
show the corresponding noiseless costs, and red and black vertical lines mark
the fitted and noiseless minima, respectively. The dotted grey curve in
\textbf{(b)} is the noiseless cost shifted by the fitted constant
\(\Delta C\), showing that the directional shape is preserved despite the
device-dependent offset. Error bars indicate one standard error from shot noise. }
\label{fig:shot-noise-parabolas}
\end{figure*}

\label{app:quantum-algorithm-details}

Within a physical time step, the current parameters and Trotterized evolution
are held fixed, and we sample the effective cost function
\(C_{\vec\varphi_t}(\vec\varphi)\) defined in
Eq.~\eqref{eq:eff-cost}. Any quantum measurement produces noisy estimates,
which can make the location of the minimum difficult to resolve, particularly
under a limited shot budget and in the presence of hardware noise.

A standard approach would be to estimate a stochastic gradient using
simultaneous perturbation stochastic approximation (SPSA). In the present
setting, however, we found that these updates were not sufficiently stable:
the gradient is inferred from the difference between two noisy cost
measurements and can therefore be dominated by statistical or device noise.
Instead, we exploit the smooth local structure of the cost landscape. We
sample one-dimensional cuts through parameter space and approximate the cost
along each cut by a quadratic function.

More precisely, for a coordinate direction \(\mathbf e_i\) and a reference
point \(\vec\varphi_0\), we fit
\begin{equation}
  C_i(\tau)
  \equiv
  C_{\vec\varphi_t}
  \bigl(\vec\varphi_0+\tau\mathbf e_i\bigr)
  \simeq
  a_i(\tau-\tau_i^\star)^2+c_i.
  \label{eq:line-parabola}
\end{equation}

The central advantage of this approach is that the minimum can be inferred
from measurements taken away from the minimum itself. Close to the minimum,
the cost is nearly flat, and differences between nearby measurements may be
smaller than the shot and hardware noise. Further along the same cut, the
quadratic increase of the cost produces a larger signal. Provided that the
sampled interval remains within the regime where the quadratic approximation
is valid, these higher-signal points constrain both the curvature and the
position of the minimum. The fit can therefore locate a minimum even when cost
differences in its immediate vicinity are not directly resolvable.

Figure~\ref{fig:shot-noise-parabolas} illustrates this behaviour. On both
devices, the individual measurements are noisy, but their variation along the
parameter cut retains an approximately quadratic form, and the fitted minima
remain close to those of the noiseless cost. The IBM measurements additionally contain a substantial vertical offset arising from device noise rather than shot noise. Over the sampled parameter cut, this offset is approximately constant and therefore shifts the measured cost without substantially changing its curvature or the location of its minimum.

The fitted minimum defines a candidate update, which is not accepted
automatically. We accept a fit only when it describes a well-resolved and
physically plausible minimum. In particular, we require positive curvature,
a minimum within the trusted parameter range, and sufficient agreement
between the quadratic model and the measured data. If the fit is accepted,
the corresponding parameter is moved to the fitted minimum and the procedure
is repeated about the updated point. If it is rejected, the parameters remain
unchanged and the protocol continues with further directional sampling. The
precise direction-selection strategy and acceptance criteria were adapted to
the different shot budgets available on the two devices, as described below.

To quantify the agreement between the quadratic model and the sampled cost,
we define the residuals
\(r_{ij}=C_i(\tau_j)-C_i^{\mathrm{fit}}(\tau_j)\) and use
\begin{align}
R_i^2
&=
1-\frac{\sum_j r_{ij}^2}
        {\sum_j [C_i(\tau_j)-\overline{C}_i]^2},
\\
\hat{\sigma}_i
&=
\sqrt{\frac{\sum_j r_{ij}^2}{N_i-3}},
\end{align}
where \(N_i\) is the number of sampled points along the cut.
The coefficient \(R_i^2\) measures how well the quadratic model captures the
variation of the cost along the cut, while \(\hat{\sigma}_i\) measures the
absolute residual scatter in units of the cost. These diagnostics complement
the geometric requirements \(a_i>0\) and a finite fitted minimum
\(\tau_i^\star\) within the trusted parameter interval. The thresholds and
additional acceptance conditions differed between the two hardware
implementations.

\paragraph{Quantinuum H2-2 protocol.} \label{sec:diamond}
The Quantinuum implementation was designed to exploit the high-fidelity,
fully connected trapped-ion architecture and its support for mid-circuit
measurement and reset. In the sequential overlap circuit, each pair of
qubits is reset after its destructive swap measurement and reused in the
subsequent contraction. This qubit recycling reduces the circuit width
required to evaluate the local overlaps.

Because the wall-clock cost of repeated trapped-ion execution is appreciable,
we designed the optimization to extract as much information as possible from
a restricted shot budget. Each cost evaluation used \(500\) shots. Rather
than densely scanning all six parameter directions, we first probed candidate
directions at the paired displacements \(\tau=\pm\delta\). The difference
\[
g_i=C_i(+\delta)-C_i(-\delta)
\]
provides a noise-robust indication that the minimum is displaced from the
current point. Directions were tested sequentially, without replacement,
until a prescribed number satisfied
\[
|g_i|>\kappa_i\sigma_{\mathrm{nom}},
\]
where \(\sigma_{\mathrm{nom}}\) is the nominal statistical scale and
\(\kappa_i\) is a step-dependent significance threshold. Among these
candidates, the direction with the largest \(|g_i|\) was selected for a
denser scan.

The selected cut was sampled at between six and ten points over the fitting interval, including the screening measurements, and fitted to Eq.~\eqref{eq:line-parabola}. An accepted fit displaced
the corresponding parameter to the fitted minimum, after which the direction
screening was repeated about the new point. If the fit was rejected, the
parameters remained unchanged and a new screening pass was initiated. We
allowed more such passes at later physical times, where maintaining an
accurate parameterization generally required greater optimization effort. In
one step for which repeated sampling of the same direction revealed clear
non-quadratic structure, we used a fourth-order fit to the combined samples;
this exceptional update is identified explicitly in the protocol record.

\paragraph*{Higher-budget finite-shot simulations.}
To estimate how the Quantinuum results could
improve with a larger sampling budget, we repeated the six variational updates
in \(100\) independent finite-shot simulations of the ideal circuits. The
simulations used the same three warm-start states and the same quadratic-fit
optimization strategy as the hardware run, but with denser sampling. All six
parameter directions were screened, each selected cut contained \(12\)
sampled points, and up to five fit-and-update attempts were allowed at each
physical time step. The purity and fidelity circuits were evaluated with
\(2000\) and \(4000\) shots, respectively, four times the corresponding
per-circuit sampling used in the H2-2 protocol. The statistical acceptance
thresholds were rescaled according to the expected \(1/\sqrt{N}\) reduction
of shot noise. This resulted in \(646\pm50\) sampled cost values per simulated
trajectory on average.

For every simulated trajectory, we computed the local and global Loschmidt
rates from the resulting variational parameters and recorded the accepted
cost after each update. The diamonds in Fig.~\ref{fig:hardware-trajectories} show the pointwise medians of these
quantities over the \(100\) simulations. No device-noise model was included;
the comparison therefore isolates the expected improvement from increasing
the sampling budget and optimization effort.

\paragraph{IBM protocol.}
The larger shot budget available for the IBM experiment allowed us to use a
more systematic direction-selection strategy. Each constituent circuit was
evaluated with \(5000\) shots. At every optimization attempt, we first sampled
ten displacements in each of the six coordinate directions over
\(-0.5\leq\tau\leq0.5\). The directions were ranked according to the validity
of their fitted geometry and the predicted reduction in the cost. The three
most promising directions were then refined to twenty sampled displacements.
If none of these refined fits produced an admissible update, the remaining
three directions were refined in the same way.

For the IBM data, a quadratic fit was considered actionable when it had
positive curvature, a finite minimum satisfying
\(|\tau_i^\star|\leq0.5\), \(R^2\geq0.5\), and a predicted reduction
\(C_i(0)-C_i(\tau_i^\star)>10^{-3}\). For larger proposed displacements,
\(|\tau_i^\star|>0.25\), we imposed the stronger requirement
\(R^2\geq0.7\). If several directions passed these tests, we selected the fit
with the largest \(R^2\), using the predicted reduction as a secondary
criterion. Optimization at a physical time step was declared converged only
after all six directions had been fully sampled without producing an
actionable update.

To detect temporal drift during a scan, the cost at the reference point was
measured both before and after the directional samples. A scan was retained
only when the difference between these two measurements was compatible with
their combined statistical uncertainty. This test distinguishes a stable
parameter dependence from changes in the effective cost caused by device
drift during data acquisition.

The data supporting this work, including the sampled hardware cost
landscapes, fit diagnostics, accepted parameter trajectories, circuit
descriptions, and analysis scripts, are available from the UCL Research Data
Repository at
\href{https://doi.org/10.5522/04/33898246}{doi:10.5522/04/33898246}~\cite{LOMPSData}.
The corresponding quantum circuit and optimization code is available in
Ref.~\cite{qlomps_code}.
Raw hardware outputs and associated provenance metadata are included where
permitted by the hardware providers.

\section{The LOMPS algorithm}
\label{app:lomps}

\subsection{Uniform MPS and local reduced density matrices}
\label{app:lomps-umps}

We represent the state by a one-site uniform MPS tensor
\begin{equation}
  A=\{A^s\}_{s=1}^{d},
  \qquad
  A^s\in\mathbb C^{D\times D},
\end{equation}
where \(d\) is the local Hilbert-space dimension and \(D\) is the bond
dimension.  The tensor is kept in left-canonical form,
\begin{equation}
  \sum_{s=1}^{d} A^{s\dagger}A^s=I_D.
  \label{eq:lomps-draft-left-canonical}
\end{equation}
It is useful to combine the physical components into the matrix
\begin{equation}
  W(A)=
  \begin{pmatrix}
    A^1\\[-0.2em]
    \vdots\\[-0.2em]
    A^d
  \end{pmatrix}
  \in\mathbb C^{dD\times D}.
  \label{eq:lomps-draft-stacked-tensor}
\end{equation}
Equation~\eqref{eq:lomps-draft-left-canonical} is then simply
\(W^\dagger W=I_D\).  The set of matrices satisfying this constraint is the
complex Stiefel manifold of \(dD\times D\) isometries.

The transfer map associated with \(A\) is
\begin{equation}
  \mathcal E_A(X)=\sum_s A^s X A^{s\dagger}.
  \label{eq:lomps-draft-transfer-map}
\end{equation}
Left-canonical form makes this map trace preserving.  We assume that the MPS
is injective, so that the normalized positive right fixed point is unique:
\begin{equation}
  \mathcal E_A(r_A)=r_A,
  \qquad
  \operatorname{Tr}r_A=1.
  \label{eq:lomps-draft-right-fixed-point}
\end{equation}
The corresponding left fixed point is the identity.  Thus the two infinite
environments surrounding a finite block reduce to \(I_D\) and \(r_A\).

For the physical string \(\boldsymbol{s}=(s_1,\ldots,s_L)\), define
\begin{equation}
  M_{\boldsymbol{s}}(A)=A^{s_1}A^{s_2}\cdots A^{s_L}.
\end{equation}
The \(L\)-site reduced density matrix of the infinite MPS is then
\begin{equation}
  \rho_L(A)_{\boldsymbol{s},\boldsymbol{t}}
  =
  \operatorname{Tr}\!\left[
    M_{\boldsymbol{s}}(A)\,r_A\,
    M_{\boldsymbol{t}}(A)^\dagger
  \right].
  \label{eq:lomps-draft-rdm}
\end{equation}
This finite object is the only part of the variational state that enters the
LOMPS cost function.

\subsection{Trotterized local target}
\label{app:lomps-trotter-target}

Consider a translation-invariant nearest-neighbour Hamiltonian
\begin{equation}
  H=\sum_j h_{j,j+1}=H_{\mathrm A}+H_{\mathrm B},
\end{equation}
where \(H_{\mathrm A}\) and \(H_{\mathrm B}\) contain the two sets of
non-overlapping bonds.  Let
\begin{equation}
  U_{\mathrm A}(\tau)=\prod_{j\in\mathrm A}e^{-i\tau h_{j,j+1}},
  \qquad
  U_{\mathrm B}(\tau)=\prod_{j\in\mathrm B}e^{-i\tau h_{j,j+1}}.
\end{equation}
We approximate one time step by the second-order brick-wall circuit
\begin{equation}
  U_{\Delta t}^{(\mathrm A)}
  =
  U_{\mathrm A}(\Delta t/2)
  U_{\mathrm B}(\Delta t)
  U_{\mathrm A}(\Delta t/2).
  \label{eq:lomps-draft-strang}
\end{equation}
A translation by one site interchanges the two bond sets and gives the
corresponding \(\mathrm B\)-origin pattern.

Because the circuit has a finite light cone, the updated \(L\)-site reduced density matrix can be
constructed from a larger but still finite reduced density matrix of the current MPS.  For even
\(L\), we use an \(L+4\)-site reduced density matrix and retain the central block after the
brick-wall step,
\begin{equation}
  \widetilde\rho_{L,n}
  =
  \operatorname{Tr}_{2|2}\!\left[
    U_{\Delta t}^{(\mathrm A)}
    \rho_{L+4}(A_n)
    U_{\Delta t}^{(\mathrm A)\dagger}
  \right].
  \label{eq:lomps-draft-even-target}
\end{equation}
Here \(\operatorname{Tr}_{p|q}\) means that \(p\) buffer sites are traced out
on the left and \(q\) on the right.

For odd \(L\), the observation window has two inequivalent placements relative
to the two bond layers.  We therefore evolve an \(L+5\)-site reduced density matrix and form the
two translated reductions
\begin{align}
  \widetilde\rho_{L,n}^{(\mathrm A)}
  &={}
  \operatorname{Tr}_{2|3}\!\left[
    U_{\Delta t}^{(\mathrm A)}
    \rho_{L+5}(A_n)
    U_{\Delta t}^{(\mathrm A)\dagger}
  \right],
  \\
  \widetilde\rho_{L,n}^{(\mathrm B)}
  &={}
  \operatorname{Tr}_{3|2}\!\left[
    U_{\Delta t}^{(\mathrm A)}
    \rho_{L+5}(A_n)
    U_{\Delta t}^{(\mathrm A)\dagger}
  \right].
  \label{eq:lomps-draft-odd-targets}
\end{align}
Relative to the retained window, these are the \(\mathrm A\)- and
\(\mathrm B\)-origin brick-wall patterns.  The odd-\(L\) target is their equal
mixture,
\begin{equation}
  \widetilde\rho_{L,n}
  =\frac12\left(
    \widetilde\rho_{L,n}^{(\mathrm A)}
    +\widetilde\rho_{L,n}^{(\mathrm B)}
  \right).
  \label{eq:lomps-draft-odd-average}
\end{equation}
This is an average of density matrices, not a coherent superposition.  It
prevents the one-site translation-invariant variational ansatz from selecting
one of the two Trotter origins.

\subsection{Representability and energy conservation}
\label{app:lomps-representability}

At physical time step \(n\), LOMPS holds the target
\(\widetilde\rho_{L,n}\) fixed and minimizes
\begin{equation}
  C_n(A)
  =
  \frac12
  \left\|\rho_L(A)-\widetilde\rho_{L,n}\right\|_F^2
  \label{eq:lomps-draft-cost}
\end{equation}
over left-canonical uniform MPS tensors of the chosen bond dimension.  The
previous tensor \(A_n\) provides the natural initial guess.  Let \(W_\star\)
be the matrix with the smallest cost found during this minimization, and
denote its \(d\) consecutive \(D\times D\) row blocks by \(W_\star^s\).  The
next MPS tensor is simply
\begin{equation}
  A_{n+1}^s=W_\star^s,
  \qquad s=1,\ldots,d.
  \label{eq:lomps-draft-update}
\end{equation}
Thus \(A_{n+1}\) is obtained after an iterative nonlinear optimization.

\subsubsection{Parameter count.}
A normalized density matrix on \(L\) sites has \(d^{2L}-1\) real parameters.
For a one-site translation-invariant state, equality of its left and right
\((L-1)\)-site marginals imposes \(d^{2L-2}-1\) independent real constraints.
The resulting local data therefore have
\begin{equation}
  N_{\mathrm{TI}}(d,L)=d^{2L}-d^{2L-2}
  \label{eq:lomps-draft-ti-dimension}
\end{equation}
real degrees of freedom.  On the variational side, the left-canonical Stiefel
manifold has real dimension \((2d-1)D^2\).  After quotienting by virtual
unitary conjugation and the common tensor phase, the physical uniform-MPS
manifold has dimension
\begin{equation}
  N_{\mathrm{MPS}}(d,D)=2(d-1)D^2.
  \label{eq:lomps-draft-mps-dimension}
\end{equation}
Consequently, a necessary generic parameter-counting condition for
representing arbitrary translation-invariant \(L\)-site data is
\begin{equation}
  2(d-1)D^2\geq d^{2L}-d^{2L-2}.
  \label{eq:lomps-draft-representability-count}
\end{equation}
For qubits, this gives
\(D\geq\lceil\sqrt{3\,4^{L-1}/2}\rceil\), motivating the scaling
\(D\sim2^L\).  This dimension count is necessary but not sufficient: the local
reduced density matrix map is nonlinear, and satisfying Eq.~\eqref{eq:lomps-draft-representability-count}
does not by itself guarantee that a particular target is representable. However, when performing the algorithm, the sufficiency of the bond dimension is automatically checked via the residual cost.

The parameter count distinguishes two regimes. When
\begin{equation}
2(d-1)D^2
\geq
d^{2L}-d^{2L-2},
\end{equation}
the MPS manifold has at least as many physical parameters as a generic
translation-invariant \(L\)-site reduced density matrix. We refer to this as
the locally expressive regime. All classical LOMPS results presented
in this work are obtained in this regime. When
\begin{equation}
2(d-1)D^2
<
d^{2L}-d^{2L-2},
\end{equation}
the variational manifold is lower dimensional than the space of generic local
data. A generic \(L\)-site reduced density matrix can then no longer be
represented exactly, and the projection cost contains an intrinsic
representation error in addition to the finite-window error. This compressed
regime may substantially reduce the computational resources, but its accuracy
and dynamics are left for future work.

\subsubsection{Energy conservation.}
For $L\geq 2$, an exact projection preserves the energy density of any
translation-invariant two-local Hamiltonian. More generally, let
$\widetilde{\rho}_{L,n}$ denote the $L$-site target obtained after the
time-evolution step, and define the projection residual

\begin{equation}
\epsilon_{L,n}
=
\left\|
\rho_L(A_{n+1})-\widetilde{\rho}_{L,n}
\right\|_F
=
\sqrt{2C_n(A_{n+1})}.
\end{equation}

For any operator $o$ supported on $r\leq L$ consecutive sites,
Hilbert--Schmidt Cauchy--Schwarz gives
\begin{align}
&\left|
\operatorname{Tr}\!\left[
o\left(
\rho_r(A_{n+1})-\widetilde{\rho}_{r,n}
\right)
\right]
\right| \\
=&
\left|
\operatorname{Tr}\!\left[
\left(o\otimes I_{d^{L-r}}\right)
\left(
\rho_L(A_{n+1})-\widetilde{\rho}_{L,n}
\right)
\right]
\right|
\nonumber\\
\leq &
d^{(L-r)/2}\,
\|o\|_F\,\epsilon_{L,n},
\label{eq:lomps-draft-local-observable-bound}
\end{align}
where $\widetilde{\rho}_{r,n}$ is the corresponding reduction of the target.
Consequently, for a translation-invariant Hamiltonian with local density
$h$ supported on $r\leq L$ sites, the projection-induced change in energy
density obeys
\begin{equation}
\left|
e(A_{n+1})-\widetilde e_n
\right|
\leq
d^{(L-r)/2}\,
\|h\|_F\,\epsilon_{L,n}.
\end{equation}
If the target is exactly representable within the variational manifold,
then the minimum residual vanishes and the projection preserves its energy
density exactly. Any remaining energy drift arises from the approximation
used to construct the time-evolved target, such as finite-step Trotter
error, rather than from the projection itself.

\subsection{Continuous-time limit}
\label{app:lomps-continuous-time}

Let \(\lambda^i\) be real local coordinates on the variational manifold and
let \(\rho_L(\boldsymbol\lambda)\) be the corresponding reduced density
matrix. Over an infinitesimal time step, the exact local target is
\begin{align}
  \rho_L^{\mathrm{ex}}(t+\dd t)
  &={}
  \rho_L(\boldsymbol\lambda(t))
  +\dd t\,\dot\rho_L^{\mathrm{ex}}
  +O(\dd t^2),
  \nonumber\\
  \dot\rho_L^{\mathrm{ex}}
  &={}-i\,\Tr_{\bar L}[H,\rho].
  \label{eq:lomps-exact-local-derivative}
\end{align}
The variational reduced density matrix at the end of the same step is
\begin{equation}
  \rho_L(\boldsymbol\lambda(t+\dd t))
  =
  \rho_L(\boldsymbol\lambda(t))
  +\dd t\,\partial_j\rho_L\,\dot\lambda^j
  +O(\dd t^2).
  \label{eq:lomps-variational-local-derivative}
\end{equation}
Substitution into the local Hilbert--Schmidt cost gives, to leading
nonvanishing order,
\begin{equation}
  C_L
  =
  \frac{\dd t^2}{2}
  \left\|
    \partial_j\rho_L\,\dot\lambda^j
    -\dot\rho_L^{\mathrm{ex}}
  \right\|_{\mathrm{HS}}^2
  +O(\dd t^3).
  \label{eq:lomps-continuous-cost}
\end{equation}
Stationarity with respect to each real velocity \(\dot\lambda^i\) therefore
yields
\begin{align}
  S_{ij}\dot\lambda^j&=b_i,
  \nonumber\\
  S_{ij}
  &={}
  \operatorname{Re}\Tr_L(
    \partial_i\rho_L\,\partial_j\rho_L
  ),
  \nonumber\\
  b_i
  &={}
  \operatorname{Re}\Tr_L(
    \partial_i\rho_L\,\dot\rho_L^{\mathrm{ex}}
  ).
  \label{eq:lomps-continuous-normal-equations}
\end{align}
Because \(\partial_i\rho_L\) is Hermitian and
\(\Tr_{\bar L}[H,\rho]\) is anti-Hermitian, the trace
\(\Tr_L[\partial_i\rho_L\Tr_{\bar L}[H,\rho]]\) is purely imaginary. Hence
the driving term can equivalently be written as
\begin{equation}
  b_i
  =-i\,\Tr_L\!\left[
    \partial_i\rho_L\Tr_{\bar L}[H,\rho]
  \right]
  =\operatorname{Im}\Tr_L\!\left[
    \partial_i\rho_L\Tr_{\bar L}[H,\rho]
  \right].
  \label{eq:lomps-continuous-driving-term}
\end{equation}
Equation~\eqref{eq:lomps-continuous-normal-equations} is the local
reduced-density-matrix analogue of the TDVP normal equation.

\subsection{Projection to the variational manifold}

The cost in Eq.~\eqref{eq:lomps-draft-cost} defines the LOMPS projection, but does not prescribe a particular optimizer. While a range of methods could be used to find the variational tensor $A$, we here use a regularized
Gauss--Newton method of Levenberg--Marquardt type. The numerical implementation used for the classical results is provided in the public LOMPS repository~\cite{lomps_code}.

During each physical time step, the target density matrix is held fixed while
an iterative optimization moves the MPS progressively closer to the minimum
of the cost function. At the current tensor $A$, we approximate the change of the local reduced
density matrix to first order,
{\small 
\begin{equation}
\rho_L(A+\delta A)
\simeq
\rho_L(A)+\mathcal J_A(\delta A),
\end{equation}
}
where $\mathcal J_A$ is the Jacobian of the map from the MPS tensor to its
local reduced density matrix.

The tangent displacement is chosen by minimizing the squared distance of this
linear approximation from the target. This inverse problem can be poorly
conditioned because the local reduced density matrix may be only weakly
sensitive to some variations of the MPS tensor. Without regularization, such
directions can produce unnecessarily large and unreliable tensor updates. We
therefore penalize the size of the displacement and solve
\begin{equation}
\label{eq:lomps-lm-overview}
\begin{aligned}
\delta A_\star
=
\underset{\delta A}{\arg\min}\,
\Bigg[
&\frac12
\left\|
\rho_L(A)+\mathcal J_A(\delta A)-\widetilde{\rho}_{L,n}
\right\|_F^2
\\
&+\frac{\lambda}{2}\|\delta A\|^2
\Bigg].
\end{aligned}
\end{equation}
The regularization parameter $\lambda$ suppresses poorly constrained
directions and stabilizes the update. The resulting trial tensor is then
retracted to the MPS manifold and accepted if it reduces the actual cost.
This procedure is repeated until the residual is
sufficiently small or no further improvement is obtained. The following
subsections give the explicit tangent coordinates, Jacobian, and retraction
used to implement these steps.

\subsubsection{Tangent coordinates and the Jacobian}
\label{app:lomps-jacobian}

We now express the Jacobian $\mathcal J_A$ appearing in
Eq.~\eqref{eq:lomps-lm-overview} in explicit coordinates and provide the
tensor and fixed-point derivatives from which it is constructed. Throughout
this subsection, $A$ denotes the current MPS tensor and $W$ its
left-canonical matrix.

Complete the columns of $W$ to an orthonormal basis, and collect the
additional columns in $W_\perp$, so that
\begin{equation}
  W^\dagger W_\perp=0,
  \qquad
  W_\perp^\dagger W_\perp=I.
\end{equation}
To avoid varying redundant gauge coordinates, we use the standard left-gauge
condition
\begin{equation}
  W^\dagger\delta W=0.
  \label{eq:lomps-draft-left-gauge}
\end{equation}
Choose matrices \(\{E_a\}\) of size \((d-1)D\times D\) that are orthonormal
with respect to the real inner product
\(\operatorname{Re}\operatorname{Tr}(E_a^\dagger E_b)=\delta_{ab}\), and
define
\begin{equation}
  H_a=W_\perp E_a.
  \label{eq:lomps-draft-tangent-basis}
\end{equation}
There are \(p=2(d-1)D^2\) such real directions. A real coordinate vector
\(\boldsymbol\xi\in\mathbb R^p\) determines the tangent variation
\begin{equation}
  \delta W(\boldsymbol\xi)
  =\sum_{a=1}^{p}\xi_a H_a
  =W_\perp X(\boldsymbol\xi).
  \label{eq:lomps-draft-coordinate-step}
\end{equation}
The coefficients \(\xi_a\) are therefore coordinates of an MPS-tensor update,
not coordinates of the density matrix itself.

To define the residual without introducing redundant complex coordinates,
choose a Hermitian operator basis \(\{F_\mu\}\) on the \(L\)-site Hilbert
space, normalized as
\begin{equation}
  \operatorname{Tr}(F_\mu F_\nu)=\delta_{\mu\nu}.
\end{equation}
The real residual components are
\begin{equation}
  e_\mu
  =
  \operatorname{Tr}\!\left[
    F_\mu\bigl(\rho_L(A)-\widetilde\rho_{L,n}\bigr)
  \right],
  \qquad
  \mu=1,\ldots,d^{2L}.
  \label{eq:lomps-draft-residual-components}
\end{equation}
This choice preserves the Hilbert--Schmidt norm, so
\(C_n(A)=\frac12\sum_\mu e_\mu^2\).

Let \(B_a^s\) denote the \(s\)th \(D\times D\) row block of \(H_a\). These
blocks form the MPS-tensor variation
\(B_a=\{B_a^s\}_{s=1}^d\). The real Jacobian is
\begin{equation}
  J_{\mu a}
  =
  \operatorname{Tr}\!\left[
    F_\mu\,D\rho_L[A][B_a]
  \right].
  \label{eq:lomps-draft-jacobian}
\end{equation}
Equivalently, \(J_{\mu a}\) is the derivative of residual component
\(e_\mu\) when the tensor is displaced along \(H_a\).

The derivative in
Eq.~\eqref{eq:lomps-draft-jacobian} is computed as follows. For a tensor variation
\(B=\{B^s\}\), the transfer-map variation is
\begin{equation}
  D\mathcal E_A[B](X)
  =
  \sum_s\left(
    B^s X A^{s\dagger}+A^s X B^{s\dagger}
  \right).
  \label{eq:lomps-draft-transfer-derivative}
\end{equation}
The corresponding change \(\delta r\) of the right fixed point obeys
\begin{equation}
  (I-\mathcal E_A)\delta r
  =D\mathcal E_A[B](r_A),
  \qquad
  \operatorname{Tr}\delta r=0.
  \label{eq:lomps-draft-fixed-point-response}
\end{equation}
Since \(I-\mathcal E_A\) annihilates the fixed point \(r_A\), the first
equation alone does not determine \(\delta r\) uniquely. The trace condition
fixes the remaining freedom and preserves the normalization of \(r_A\).

The variation of an MPS product is
\begin{equation}
  \delta M_{\boldsymbol{s}}[B]
  =
  \sum_{j=1}^{L}
  A^{s_1}\cdots A^{s_{j-1}}B^{s_j}
  A^{s_{j+1}}\cdots A^{s_L}.
  \label{eq:lomps-draft-product-derivative}
\end{equation}
Using Eqs.~\eqref{eq:lomps-draft-fixed-point-response} and
\eqref{eq:lomps-draft-product-derivative}, the derivative is
\begin{align}
  D\rho_L[A][B]_{\boldsymbol{s},\boldsymbol{t}}
  ={}&
  \operatorname{Tr}\!\left[
    \delta M_{\boldsymbol{s}}[B]\,r_A
    M_{\boldsymbol{t}}^\dagger
  \right]
  \nonumber\\
  &+\operatorname{Tr}\!\left[
    M_{\boldsymbol{s}}\,\delta r\,
    M_{\boldsymbol{t}}^\dagger
  \right]
  \nonumber\\
  &+\operatorname{Tr}\!\left[
    M_{\boldsymbol{s}}\,r_A\,
    \delta M_{\boldsymbol{t}}[B]^\dagger
  \right].
  \label{eq:lomps-draft-rdm-derivative}
\end{align}
The three terms respectively describe variations of the ket tensors, the
infinite environment, and the bra tensors.

\subsubsection{Regularized Gauss--Newton step and retraction}
\label{app:lomps-lm-step}

For a small tangent displacement, the residual is approximated by
\begin{equation}
  e_\mu\bigl(W+\delta W(\boldsymbol\xi)\bigr)
  \simeq
  e_\mu+\sum_a J_{\mu a}\xi_a.
  \label{eq:lomps-draft-linearized-residual}
\end{equation}
The Levenberg--Marquardt step \(\boldsymbol\xi\) minimizes the regularized
linear model
\begin{equation}
  \frac12\sum_\mu
  \left(e_\mu+\sum_aJ_{\mu a}\xi_a\right)^2
  +\frac{\lambda}{2}\sum_a\xi_a^2.
  \label{eq:lomps-draft-lm-model}
\end{equation}
The coefficients therefore satisfy
\begin{equation}
  \sum_b\left[
    \sum_\mu J_{\mu a}J_{\mu b}
    +\lambda\delta_{ab}
  \right]\xi_b
  =
  -\sum_\mu J_{\mu a}e_\mu.
  \label{eq:lomps-draft-lm-equation}
\end{equation}
The limit \(\lambda\rightarrow0\) gives the Gauss--Newton step. Positive
\(\lambda\) suppresses directions to which the local reduced density matrix is weakly
sensitive and makes the linear system nonsingular.  A maximum allowed value
of \(\|\boldsymbol\xi\|_2\) can additionally bound the distance over which
the linearized model is trusted.

Solving Eq.~\eqref{eq:lomps-draft-lm-equation} gives the tangent direction
\begin{equation}
  \delta W=\sum_a\xi_aH_a.
  \label{eq:lomps-draft-lm-tangent-step}
\end{equation}
The linear displacement \(W+\delta W\) does not satisfy the
left-canonical constraint exactly.  We therefore use the polar retraction, a
smooth projection back to the Stiefel manifold,
\begin{equation}
  R_W(Y)
  =
  (W+Y)\left[(W+Y)^\dagger(W+Y)\right]^{-1/2}
  \label{eq:lomps-draft-polar-retraction}
\end{equation}
and test the candidate
\begin{equation}
  W_{\mathrm{trial}}
  =R_W\!\left(\alpha\delta W\right).
  \label{eq:lomps-draft-inner-update}
\end{equation}
A backtracking line search reduces \(\alpha\) until the actual cost decreases
sufficiently. The damping \(\lambda\) is increased when a trial direction
or line search is unsuccessful and decreased after a good step.  After an
accepted update, we set \(W\leftarrow W_{\mathrm{trial}}\) and recompute the
fixed point, reduced density matrix, tangent basis, and Jacobian at the new
left-canonical matrix.

The optimization terminates when the local reduced density matrix residual is sufficiently small
or when further tangent updates no longer improve the cost.

\bibliographystyle{apsrev4-1}
\bibliography{references}

\end{document}